\documentclass[aps,showpacs,preprint,pre,superscriptaddress]{revtex4}
\usepackage{graphicx,epsfig}
\input amssym.tex

\begin{document}
\title{Statistics of 2$^{+}$ Levels in Even--Even Nuclei}
\author{A. Y. Abul--Magd}
\affiliation{Faculty of Science, Zagazig University, Zagazig, Egypt}
\author{H.L. Harney}
\affiliation{Max-Planck-Institut f\"{u}r Kernphysik, Heidelberg}
\author{M.H. Simbel}
\affiliation{Faculty of Science, Zagazig University, Zagazig, Egypt}
\author{H.A. Weidenm{\"u}ller}
\affiliation{Max-Planck-Institut f\"{u}r Kernphysik, Heidelberg}
\date{October 2003}

\begin{abstract}
Using all the available empirical information, we analyze the spacing
distributions of low--lying 2$^{+}$ levels of even--even nuclei. To obtain
statistically relevant samples, the nuclei are grouped into classes defined
by the ratio $R_{4/2}$ of the exitation energies of the first 4$^{+}$ and 2$%
^{+}$ levels. This ratio serves as a measure of collectivity in nuclei. With
the help of Bayesian inference, we determine the chaoticity parameter for
each class. This parameter is found to vary strongly with $R_{4/2}$ and
takes particularly small values in nuclei that have one of the dynamical
symmetries of the interacting Boson model.
\end{abstract}
\maketitle

\section{Motivation and Purpose} 
\label{sect:1}
During the past decades, a vast amount of
nuclear spectroscopic data has been accumulated. Level schemes involving
tens and sometimes hundreds of levels with reliably known values of spin and
parity are now available for hundreds of nuclei (see Ref.~\cite{NDS}). The
wealth of published spectroscopic data allows for an extensive study of the
level statistics of nuclei at low excitation energies. In this paper we
report on the statistical analysis of low--lying states with spin and parity
2$^{+}$.

The interest in such a study derives from the success of random--matrix
theory (RMT) in describing the spectral properties of nuclear levels
(actually: resonances) near neutron threshold and proton threshold~\cite
{haq,watson}. Careful analysis has shown that the spectral fluctuation
properties of these resonances are in very good agreement with the
predictions of the Gaussian orthogonal ensemble (GOE) of random matrices.
This statement applies, in particular, to the nearest--neighbor spacing
(NNS) distribution which is well approximated by Wigner's surmise~\cite
{wigner} 
\begin{equation}
p_{{\rm {W}}}(s) = \frac{\pi}{2} s \exp \left( -\frac{\pi}{4} s^{2} \right)
\ .  \label{1}
\end{equation}
Here, $s$ is the NNS in units of the mean spacing. In view of the conjecture
by Bohigas, Giannoni and Schmit~\cite{bohigas}, the agreement between the
spectral fluctuation properties of the resonances and the GOE predictions
was taken as an indication of chaotic motion in medium--weight and heavy
nuclei near neutron threshold. Interest then turned to the ground--state
domain. Here, integrable models often successfully describe the
spectroscopic data, and one would, therefore, expect the spectral
fluctuation properties to be close to those predicted for regular systems.
For such systems, the NNS distribution is generically given by the Poisson
distribution, 
\begin{equation}
p_{{\rm {P}}}(s) = \exp \left( -s \right) \ .  \label{2}
\end{equation}
A statistical analysis requires complete (few or no missing levels) and pure
(few or no unknown spin--parities) level schemes. Some 15 years ago,
complete and pure level schemes were available for only a limited number of
nuclei (see, e.g., Refs.~\cite{egidy,egidy1}). The work of Ref.~\cite{aw}
then suggested that the NNS distribution of low--lying nuclear levels lies
between the Wigner and the Poisson distributions. The evidence presented in
Ref.~\cite{aw} has since become an established fact through the work in
Refs.~\cite{mitchell,raman,shriner,as,garrett,enders,shriner1}.

The wealth of spectroscopic data now available in the Nuclear Data tables~
\cite{NDS} has motivated us to investigate once again the nuclear
ground--state domain. We are able to make more definitive and precise
statements about regularity versus chaos in this domain than has been
possible so far. As in Ref.~\cite{aw}, we focus attention on 2$^{+}$ states
of select even--even nuclei. These nuclei are grouped into classes. The
classes are defined in terms of the ratio $R_{4/2}$, i.e., the ratio of the
excitation energies of the first 4$^{+}$ and the first 2$^{+}$ level in each
nucleus. We argue below that the classes define a grouping of nuclei that
have common collective behavior. The sequences of 2$^{+}$ states are
unfolded and analyzed with the help of Bayesian inference. The chaoticity
parameter $f$ defined below is determined for each class.

\section{Data Set} 
\label{sect:2}
The data on low--lying 2$^{+}$ levels of even--even nuclei
are taken from the compilation by Tilley {\it et  al.}~\cite{tilley} for
mass numbers $16$ $\leq A \leq 20$, from that of Endt~\cite{endt} for $20
\leq A \leq 44$, and from the Nuclear Data Sheets~\cite{NDS} for heavier
nuclei. We considered nuclei for which the spin--parity $J^{\pi }$
assignments of at least five consecutive 2$^{+}$-levels are unambiguous. In
cases, where the spin-parity assignments were uncertain and where the most
probable value appeared in brackets, we accepted this value. We terminated
the sequence when we arrived at a level with unassigned $J^{\pi }$, or when
an ambiguous assignment involved a 2$^{+}$ spin-parity among several
possibilities, as e.g. $J^{\pi }=(2^{+}$, $4^{+}$). We made an exception
when only one such level occurred and was followed by several unambiguously
assigned levels containing at least two 2$^{+}$ levels, provided that the
ambiguous 2$^{+}$ level is found in a similar position in the spectrum of a
neighboring nucleus. However, this situation occurred for less than 5\% of
the levels considered. In this way, we obtained 1306 levels of spin--parity 2%
$^{+}$ belonging to 169 nuclei. The composition of this ensemble is as
follows: 5 levels from each of 47 nuclei, 6 levels from each of 32 nuclei, 7
levels from each of 22 nuclei, 8 levels from each of 22 nuclei, 9 levels
from each of 16 nuclei, 10 levels from each of 14 nuclei, 11 levels from
each of 5 nuclei, 12 levels from each of 2 nuclei, and sequences of 13, 14,
15, 17, 20, 21, 24, 30, and 32 levels, each belonging to a single nucleus.

\section{Classification of Nuclei} 
\label{sect:3}
A class is defined by choosing an interval
within which the ratio 
\begin{equation}
R_{4/2} = E(4_{1}^{+}) / E(2_{1}^{+})  \label{2a}
\end{equation}
of excitation energies of the first 4$^{+}$ and the first 2$^{+}$ excited
states, must lie. The width of the intervals was taken to be 0.1 when the
total number of spacings falling into the corresponding class was about 100
or more. Otherwise, the width of the interval was increased. 
The use of the parameter (\ref{2a}) as an indicator of collective
dynamics is justified both empirically and by theoretical arguments. We
recall the arguments in turn.

(i) Casten {\it et al.}~\cite{casten} plotted $E(4_{1}^{+})$ versus $%
E(2_{1}^{+})$ for all nuclei with $38 \leq Z \leq 82$ and with $2.05 \leq
R_{4/2} \leq 3.15$. The authors found that the data fall on a straight line.
This suggests that nuclei in this wide range of $Z$--values behave like
anharmonic vibrators with nearly constant anharmonicity. As the ratio $%
R_{4/2}$ approaches the rotor limit $R_{4/2} = 3.33$, the slope of the curve
showing $E(4_{1}^{+})$ versus $E(2_{1}^{+})$ decreases within a narrow range
of $E(2_{1}^{+})$--values, asymptotically merging the rotor line of slope
3.33. In a subsequent paper~\cite{zamfir} it was found that a linear
relation between $E(4_{1}^{+})$ and $E(2_{1}^{+})$ holds for pre--collective
nuclei with $R_{4/2} < 2$. Thus, from an empirical perspective, the
dynamical structure of medium--weight and heavy nuclei can be quantified in
terms of $R_{4/2}$.

(ii) Theoretical calculations based on the IBM-1 model~\cite{iachello}
support the conclusion that $R_{4/2}$ is an appropriate measure for
collectivity in nuclei. The model has three dynamical symmetries, obtained
by constructing the chains of subgroups of the $U(6)$ group that end with
the angular momentum group $SO(3)$. The symmetries are labeled by the first
subgroup appearing in the chain which are $U(5)$, $SU(3)$, and $O(6)$
corresponding, respectively, to vibrational, rotational and $\gamma$%
--unstable nuclei. Extensive numerical calculations for the classical as
well as the quantum--mechanical IBM Hamiltonian by Alhassid {\it et al.}~
\cite{alhassid1} indeed showed a considerable reduction of the standard
measures of chaoticity when the parameters of the IBM model approach one of
the three cases just mentioned. The IBM calculation of energy levels yields
values of $R_{4/2} = 2.00$, $3.33$, and $2.50$ for the dynamical symmetries $%
U(5)$, $SU(3)$, and $O(6)$, respectively. Thus, we may expect increased
regularity of nuclei having one of these values of $R_{4/2}$.

\begin{figure}[htbp]
\centering
\rotatebox{270}{
               \includegraphics[width=9cm]{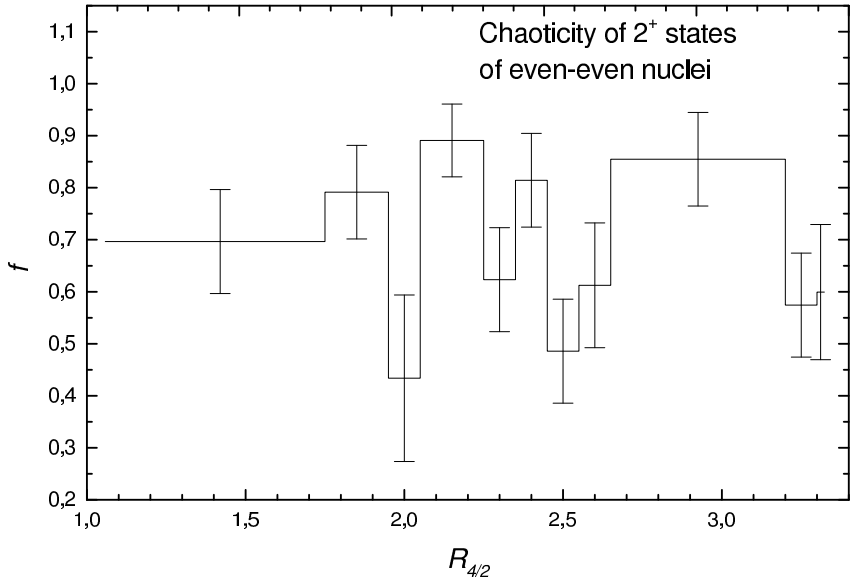}
              }
\caption{Mean value $\overline{f}$ (solid lines) and $\sigma$ (error bars)
(see Eqs.~(\ref{12})) of the chaoticity parameter for nuclei in several
classes defined in terms of $R_{4/2}$, obtained by Bayesian inference.}
 \label{fig:1} 
\end{figure}

One might expect that the chaoticity parameter also assumes small 
values for nuclei near magic numbers, where $R_{4/2}\approx 1$. 
For mass numbers in this domain, our data set is unfortunately too 
small to allow us to draw definitive conclusions, see
Fig. \ref{fig:1}.

\section{Unfolding} 
\label{sect:4}
Every sequence has to be ''unfolded'', see Ref.~\cite
{bohigas1}, to obtain a new sequence with unit mean level spacing. This is
done by fitting a theoretical expression to the number $N(E)$ of levels
below excitation energy $E$. The expression used here is the
constant--temperature formula~\cite{egidy}, 
\begin{equation}
N(E) = N_{0} + \exp \biggl( \frac{E - E_{0}}{T} \biggr) \ .  \label{13}
\end{equation}
We deal with many short sequences of levels. In this case, the unfolding
procedure introduces a bias towards the GOE. This is shown and discussed in
Ref.~\cite{ahsw} and will have to be taken into account when we discuss our
results. The three parameters $N_{0}$, $E_{0}$ and $T$ obtained for each
nucleus vary considerably with mass number. Nevertheless, all three show a
clear tendency to decrease with increasing mass number. For the effective
temperature, for example, we find, assuming a power--law dependence, the
result $T = (15 \pm 4) A^{-(0.62 \pm 0.05)}$ MeV. This result is consistent
with an analysis of the level density of nuclei in the same range of
excitation energy carried out by von Egidy {\it et al.}~\cite{egidy1}. These
authors find $T = (19 \pm 2) A^{-(0.68 \pm 0.02)}$ MeV.

\section{Method of Analysis} 
\label{sect:5}
A detailed account of our method has been given in
Ref.~\cite{ahsw}. Here we confine ourselves to the central aspects. We are
guided by the idea that the intermediate behavior of the NNS distribution of
low--lying nuclear levels does not necessarily imply that nuclei in the
vicinity of the ground state have mixed regular--chaotic dynamics. The key
ingredient of our analysis is the assumption that the deviation of the NNS
distribution of low--lying nuclear levels from the GOE statistics is caused
by the neglect of possibly existing conserved quantum numbers other than
energy, spin, and parity. A given sequence $S$ of levels can then be
represented as a superposition of $m$ independent sequences $S_{j}$ each
having fractional level density $f_{j}$, with $j=1,...,m$, and with $%
0<f_{j}\leq 1$ and $\sum_{j=1}^{m}f_{j}=1$. We assume that the NNS
distribution $p_{j}(s)$ of $S_{j}$ obeys GOE statistics. The exact NNS
distribution $p(s)$ has been given in Ref.~\cite{mehta}.
It depends on the $(m-1)$ parameters $f_{j}$, $j=1,\ldots ,m-1$. 
In \cite{as1}, this expression has been simplified by observing that
$p(s)$ is mainly determined by short-range level correlations. 
This reduces the number of parameters to unity and the proposed NNS 
distribution of the spectrum is
\begin{equation}
p(s,f)=[1-f+Q(f)\frac{\pi s}{2}]\ \exp \ [-(1-f)s-Q(f)\frac{\pi s^{2}}{4}]\, .
\label{3}
\end{equation}
Here, $f=\sum_{j=1}^{n}f_{j}^{2}$ is the mean fractional level density for
the superimposed sequences; it is the single parameter characterizing the
distribution. We determine the function $Q(f)$ from the requirement that
the expectation value of $s$ is unity, 
$\int {\rm d}s\, sp(s,f)=1$. 
This relates $Q$ to the error function. We have numerically approximated it
and obtain for $f$ in the interval of $0.1$ $\leq f\leq 0.9$ the 
parabolic relation 
\begin{equation}
Q(f)=f\left( 0.7+0.3f\right) \ .  \label{4}
\end{equation}
For a superposition of a large number $m$ of sequences, $f$ is of order $1/m$%
. In the limit of $m\rightarrow \infty $, $p(s,f)\rightarrow p(s,0)=$ $p_{%
{\rm P}}(s)$ as given by Eq.~(\ref{2}). This expresses the well--known fact
that the superposition of very many GOE sequences produces a Poisson
distribution. On the other hand, for $f\rightarrow 1$, $p(s,f)$ approaches
the Wigner distribution~(\ref{1}) expected for a single GOE. We therefore
refer to $f$ as to the chaoticity parameter. Our parameterization~(\ref{3})
is not restricted to statistically independent sequences $S_{j}$. A system
with partially broken symmetries can also be approximately represented by a
superposition of independent sequences~\cite{dembo}. In this case, the
distribution~(\ref{3}) which differs from zero at $s=0$, is not accurate for
a domain of very small spacings. The magnitude of this domain depends on the
ratio of the strength of the symmetry--breaking interaction and the mean
level spacing.

We determine the parameter $f$ by the method of Bayesian inference~\cite
{dembo}. Given a sequence of spacings ${\bf s}$ $=(s_{1},s_{2},...,s_{N})$,
the joint probability distribution $p({\bf s}|f)$ of these spacings,
conditioned by the parameter $f$, is given by 
\begin{equation}
p({\bf s}|f)=\prod_{i=1}^{N}p(s_{i},f).  \label{5}
\end{equation}
Eq.~(\ref{5}) holds if the experimental $s_{i}$ are taken to be
statistically independent. This assumption is justified as long as we
confine ourselves to the investigation of the NNS distribution. Bayes'
theorem then provides the posterior distribution 
\begin{equation}
P(f|{\bf s})=\frac{p({\bf s}|f)\mu (f)}{M({\bf s})}  \label{6}
\end{equation}
of the parameter $f$ given the events ${\bf s}$. Here, $\mu (f)$ is the
prior distribution and $M({\bf s})=\int_{0}^{1}p({\bf s}|f)\mu (f){\rm d}f$
is the normalization. The prior distribution is found from Jeffreys' rule 
\cite{jeffreys,harney}
\begin{equation}
\mu (f)\propto \biggl|\int p\left( {\bf s}\left| {}\right. f\right) \ \left[
\ \partial \ \ln p\left( {\bf s}\left| {}\right. f\right) /\ \partial f\ %
\right] ^{2}\ {\rm d}{\bf s}\ \biggr|^{1/2}\ .  \label{15}
\end{equation}
We substitute Eq.~(8) into formula~(\ref{15}), evaluate the integral
numerically and approximate the result by the polynomial 
\begin{equation}
\mu (f)=1.975-10.07f+48.96f^{2}-135.6f^{3}+205.6f^{4}-158.6f^{5}+48.63f^{6}\
.  \label{7}
\end{equation}
Even for only moderately large $N$, it is useful to write $p({\bf s}|f)$ in
the form 
\begin{equation}
p({\bf s}|f)=e^{-N\phi (f)}\,,  \label{8}
\end{equation}
where 
\begin{equation}
\phi (f)=(1-f)\langle s\rangle +\frac{\pi }{4}f(0.7+0.3f)\langle
s^{2}\rangle -\langle \ln [1-f+\frac{\pi }{2}f(0.7+0.3f)s]\rangle \ .
\label{9}
\end{equation}
Here the notation $\langle x\rangle =(1/N)\sum_{i=1}^{N}x_{i}$ has been
used. By calculating the mean values $\langle \cdots \rangle $ in Eq.~(\ref
{9}) for various spectra, one finds that the function $\phi (f)$ has a deep
minimum, say at $f=f_{0}$. One can therefore represent the numerical results
in analytical form by parametrizing $\phi $ as 
\begin{equation}
\phi (f)=A+B(f-f_{0})^{2}+C(f-f_{0})^{3}\ .  \label{10}
\end{equation}
We then obtain 
\begin{equation}
P(f|{\bf s})=c\mu (f)\exp (-N[B(f-f_{0})^{2}+C(f-f_{0})^{3}])\ ,  \label{11}
\end{equation}
where $c=e^{-NA}/M({\bf s})$ is a normalization constant. The error interval 
$\overline{f}\pm \sigma ^{1/2}$ of the chaoticity parameter is defined by
the mean value $\overline{f}$ and the variance $\sigma ^{2}$, with 
\begin{equation}
\overline{f}=\int_{0}^{1}fP(f|{\bf s})\ {\rm d}f\ \ {\rm and}\ \ \sigma
^{2}=\int_{0}^{1}(f-\overline{f})^{2}P(f|{\bf s})\ {\rm d}f\ .  \label{12}
\end{equation}

\section{Chaoticity Parameter} 
\label{sect:6}
\begin{figure}[htbp]
\centering
\rotatebox{270}{
               \includegraphics[width=10cm]{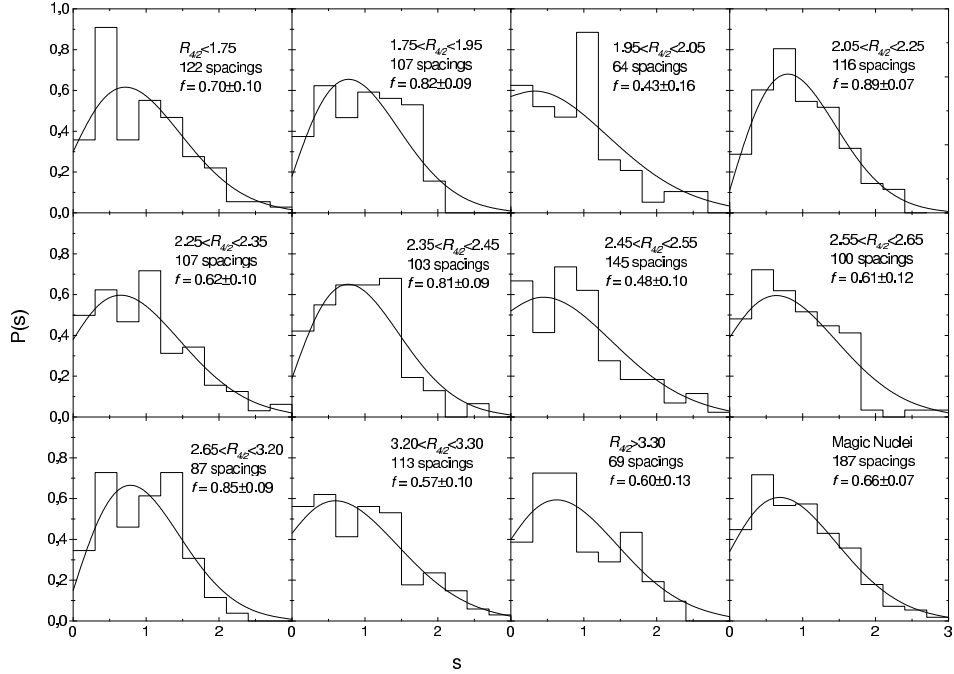}
              }
\caption{Comparison of the spacing distributions calculated from 
Eq.~(\ref{3}) using the values of $\overline{f}$ given in 
Fig. \ref{fig:1} with the histograms for
the empirical NNS distributions for nuclei in several classes defined in
terms of $R_{4/2}$.}
 \label{fig:2} 
\end{figure}

The results obtained for $\overline{f}$ and $%
\sigma$ are given in Fig. \ref{fig:1}. Figure \ref{fig:2} shows a 
comparison of the spacing
distributions conditioned by $\overline{f}$ and the histograms for each
class of nuclei. In view of the small number of spacings within each class,
the agreement seems satisfactory.

We recall that the analysis of many short sequences of levels tends to
overestimate $\overline{f}$. Therefore, we focus attention not on the
absolute values of $\overline{f}$ but on the way $\overline{f}$ changes with 
$R_{4/2}$. The graph of $\overline{f}$ against $R_{4/2}$ in Figure~1 has
deep minima at $R_{4/2}$ $= 2.0, 2.5$, and $3.3$. These values of $R_{4/2}$
are associated with the dynamical symmetries of the IBM mentioned above.
Another minimum of statistical significance occurs for $2.25 \leq R_{4/2}
\leq 2.35$. This minimum may indicate that nuclei which lie between the
limiting cases of the $U(5)$ and $O(6)$ dynamical symmetries, are relatively
regular. One may associate this region with the critical point of the $U(5)$%
--$O(6)$ shape transition in nuclei. Iachello~\cite{iachello1} has recently
shown that this transition is approximately governed by the ''critical'' $%
E(5)$ dynamical symmetry. Nuclei with $E(5)$ dynamical symmetry have $R_{4/2}
$ $= 2.2$. Experimental examples of this critical symmetry have been found
by Casten and Zamfir~\cite{casten1}.

\section{Summary} 
\label{sect:7}
With the help of a systematic analysis of the NNS
distributions for 2$^{+}$ levels of even--even nuclei, we have determined
the chaoticity parameter $f$ for nuclei at low excitation energy. While in a
single nucleus the number of states with reliable spin--parity assignments
is not sufficient for a meaningful statistical analysis, a combination of
sequences of levels taken from similar nuclei provides a sufficiently large
ensemble. As the measure of similarity we have taken the ratio $R_{4/2}$ of
the excitation energies of the lowest 4$^+$ and 2$^+$ levels in each
nucleus. As seen in Figure~1, the chaoticity parameter $\overline{f}$ is
indeed dependent on $R_{4/2}$. It has deep minima at $R_{4/2} = 2.0$, $2.5$,
and $3.3$. These minima correspond, respectively, to the $U(5)$, $SO(6)$,
and $SU(3)$ dynamical symmetries of the IBM. A further minimum may relate to
the critical $E(5)$ symmetry.


The authors thank Professor J. H\"{u}fner for useful discussions. A. Y.
A.--M. and M. H. S. acknowledge the financial support granted by
Internationales B\"{u}ro, Forschungszentrum J\"{u}lich which permitted their
stay at the Max--Planck--Institut f\"{u}r Kernphysik, Heidelberg.


\end{document}